\documentclass{epl}
\title{Non-equilibrium gelation transition\\ in a kinetic lattice gas model}
\shorttitle{Non-equilibrium Gelation Transition...}
\author{V. I. Tokar\inst{1,2} \and H. Dreyss\'e\inst{1}} 
\institute{
  \inst{1} IPCMS-GEMM, UMR 7504 CNRS, 23, rue du Loess, 
F-67037 Strasbourg Cedex, France\\
  \inst{2} Institute of Magnetism, National Academy of Sciences, 
36-b Vernadsky str., 03142 Kiev-142, Ukraine}

\begin{document}

\maketitle

\begin{abstract} 
We consider a lattice gas model which in addition to the canonical
nearest neighbor pair interatomic interaction accounts for a many-body
interaction inside atomic trios.  Interactions of this kind arise in
the coherent strained epitaxy and were recently used by us to describe
some surface phenomena. With the use of the Monte Carlo simulation we
show that in two dimensions the model at low temperature exhibits
glassy behaviour, in particular, undergoes a gelation transition.  We
argue that this model may belong to the universality class of
non-equilibrium critical phenomena which may comprise also some
off-lattice structural glass transitions, provided such a class 
exists.
\end{abstract}
It seems to be firmly established that the glass
transition which has been the subject of extensive study in recent
years (see review papers~\cite{angell,nature,ritort}) does not belong
to the category of the familiar equilibrium phase transitions but is a
purely kinetic phenomenon which occurs in some many-body far from
equilibrium systems (FFES)~\cite{angell}.  Thus, our poor understanding
of glassy phenomena~\cite{binder0,nature} can be related to the poor
development of the non-equilibrium statistical physics in comparison
with its equilibrium counterpart.

Historically, investigation of the non-equilibrium kinetics went in
parallel with the equilibrium studies~\footnote{The famous Boltzmann
equation (1872) appeared practically simultaneously with the modern
formulation of the laws of thermodynamics.} but only for systems in the
vicinity of equilibrium a systematic approach based on the linear
response theory has been developed~\cite{foerster} while in the
far from equilibrium case any technique of comparable generality
is absent.

A major difficulty in developing a unified description of the FFES is
their much larger diversity in comparison with the near-equilibrium
ones~\footnote{While near equilibrium systems in statistical terms can
be considered as linear perturbations in the vicinity of a canonical
equilibrium ensemble, the FFES comprise all other statistical
ensembles.} and that some of them are very complex, biological systems
being the standard example. So it seems unrealistic to expect that a
microscopic approach similar to LRT can be built which would be equally
applicable to all FFES. Therefore, alongside with microscopic theories
of individual FFES, phenomenological approaches are being
developed~\cite{npt,ritort} based on simple lattice models aiming at a
qualitative description of typical non-equilibrium behaviours observed
in physically different systems.  It is hoped that the number of such
models is small~\cite{npt,ritort} which would provide us with an
efficient classification of far from equilibrium phenomena similar to
the classification of the equilibrium critical phenomena according to
their universality classes~\cite{ma}.

This approach looks as a natural candidate to attack the problem of the
glass transition.  Indeed, this phenomenon occurs in such disparate
systems as metal alloys, polymer melts, biopolymer~\cite{angell},
colloidal suspensions~\cite{colloids,colloids2}, etc., that their
unified microscopic description looks hopeless. On the other hand, the
physics of the glassy phenomena is considered to be basically the same
in all substances and their theoretical interpretation is being usually
sought in a unified framework~\cite{nature}.

Presumably with this aim a number of lattice models of glassy phenomena
have been proposed~\cite{ritort}.  The main deficiency of the majority
of those models is, in our opinion, that their definitions are not
quite physical.  For example, the model of Kob and Andersen~\cite{koba}
is based on microscopic dynamics defined by a set of rules designed to
imitate the glassy behaviour. But the physical origin of these rules is
not clear, especially taking into account that the kinetics do not
change the system's energy. The recently proposed class of lattice
glass models~\cite{mezard} replaces the dynamic constraints of
ref.~\cite{koba} by similar constraints on the local atomic density. In
both cases the proposed rules are motivated by quite plausible physical
arguments. But recently it was shown~\cite{rules1,rules2} that contrary
to what one might expect on the basis of experience gained in dealing
with equilibrium critical phenomena, in the far from equilibrium case
even a seemingly innocent and commonly accepted simplification of
microscopic dynamic rules can change the universality class of the
non-equilibrium critical behaviour.

The aim of the present letter is to propose a model which is capable of
describing a glass transition and at the same time is physical in the
same sense as the canonical lattice gas model (LGM) with nearest
neighbor interatomic interactions. The latter is equivalent to the
Ising model which up to date is the most widely used model of  both
equilibrium and non-equilibrium phenomena.  In our opinion, the unique
status of the LGM/Ising model in statistical physics is largely due to
the fact that while being extremely simple and thus very convenient for
theoretical study, the model at the same time is quite physical in the
sense that from the point of view of statistical physics it might have
described a real system.  Similarly, the stochastic dynamics commonly
used in numerical simulations of FFES were shown~\cite{liggett} to be
formally fully consistent with the statistical laws, hence also
physically acceptable.

The hamiltonian of the model we are going to consider reads
\begin{equation}
\label{Hp}
H= \sum_{\bf i , \gamma}({V}{n}_{\bf i}
{n}_{\bf i + \gamma}
+ W{n}_{\bf i - \gamma}{n}_{\bf i}{n}_{\bf i + \gamma}),
\end{equation}
where $\gamma=\{\hat{e}_x,\hat{e}_y\}$ are the unit vectors of a square
lattice.  The parameters $V$ and $W$ are the pair nearest neighbor (NN)
interaction and the trio interaction parameters, respectively. This
hamiltonian can be derived as a model of strained epitaxy~\cite{us} and
is essentially equivalent to the model of ref.~\cite{model1}. Elsewhere
we have shown that this model at low surface coverage can describe the
formation of square $2\times2$ plaquettes~\cite{us}. This may serve
either as an oversimplified model of the self-assembly of quantum dots
or as a model of chemical reaction between the lattice atoms having the
4-atom molecules as its product.  Furthermore, at higher coverage the
model exhibited a surface reconstruction~\cite{us2}. All the phenomena
listed take place at $W>0$ which physically corresponds to the
positiveness of the elastic energy from which this term
originates~\cite{us}.

In the present paper we consider the case $W<0$. Physically this may be
justified in the case of strained epitaxy by considering the vacancies
instead of atoms, {\em i. e.}, making in eq. (\ref{Hp}) the substitution
$n_i\rightarrow1-n_i$. We neglect the positive next NN (NNN) pair
interaction appearing under this transformation by assuming that it is
compensated for by an attractive NNN interatomic interaction.  This
assumption is made to keep the resulting hamiltonian as simple as
possible, in the spirit of the renormalization group
approach~\cite{ma}.  We stress once again that while we are unaware of
any real system which would exactly correspond to
the hamiltonian~(\ref{Hp}), a physically acceptable interatomic interaction
potential and the lattice size misfit can be chosen so that the model
described a plausible heteroepitaxial system.  In view of possible
application of the model in strained epitaxy,~\footnote{In
ref.\ \cite{lannoo-} a long lived disordered wiggled vacancy structure
in a partially filled wetting layer was observed.  This might have been
caused by the underlying glassy kinetics.} below we consider the model
where the ``atoms'' of the hamiltonian~(\ref{Hp}) are, in fact, the
vacancies while it is the real atoms which are moving. We note,
however, that simulations also were made with $n_i$ being treated as
atoms and no qualitative differences with the behaviour described below
was found. 

To study the kinetics of the model we consider an ensemble of atoms
randomly deposited at time $t =0$ (the instant quench; see fig.\ \ref{
fig1}a).  The microscopic dynamics chosen was the activated atomic
hopping which is a natural choice for the surface
atoms~\cite{thereference}.  The parameters in eq.\ (\ref{Hp}) and the
temperature were chosen so that 
\begin{equation} 
\label{ 2}
{V}/|W|=0.75\qquad\mbox{ and}\qquad {V}/k_BT=6.  
\end{equation} 
We would like to stress that this choice does not correspond to any
exact degeneracy because we consider an accidental degeneracy to be
highly improbable in real systems.  The glassy behaviour remains
qualitatively the same if the value of $V/W$ is slightly changed as
well as if some weak additional interactions are present which do not
change the most essential qualitative feature of the hamiltonian
(\ref{Hp}): the competition between the NN repulsion due to the pair
interaction and the multiatom attraction within atomic trios. The
minimum of $H$ is unique and corresponds to a phase separated state of
vacancies and atoms.  Physically this corresponds to a partly filled
epitaxial layer.  The temperature was chosen to be high enough to allow
us to follow the evolution of the system to the very end.  At lower
temperatures the CPU time scale quickly grows because of the ``futile''
dynamics similar to that described in ref.\ \cite{novotny} when a small
number of rapidly fluctuating particles hinders major rearrangements on
broader scales.
\begin{figure}
\includegraphics[viewport = -45 300 230 710, scale = 0.55]{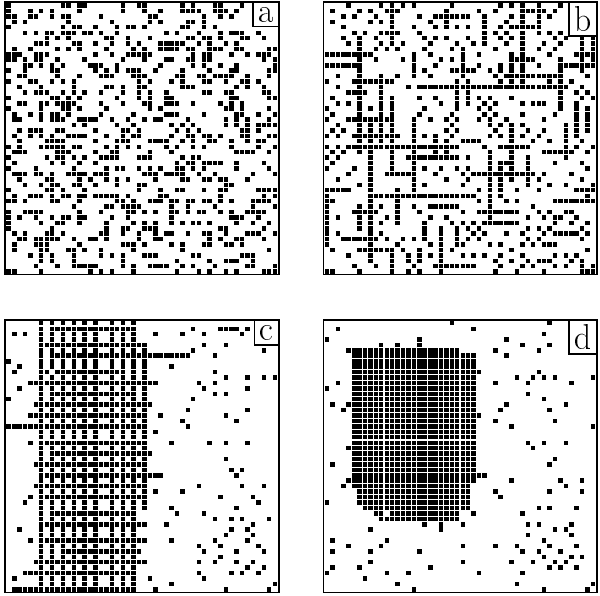}
\caption{Snapshots of the system at different times of evolution:
{\em a)} $t=0$; {\em b)} $t=5\cdot10^{-3}$; {\em c)} $t=1$; 
\mbox{{\em d)} $t=1.5$}.}
\label{ fig1}
\end{figure} 

Monte Carlo (MC) simulations were performed on finite-sized lattices of
linear dimension $L$ with periodic boundary conditions and the vacancy
concentration $c = 1/3$ which is below the percolation threshold which
for the square lattice is 0.5~\cite{stauffer}.  The time scale was
normalized so that the hopping probability of a vacancy in the atomic
bulk was equal to unity.

The evolution of the system proceeded as follows.

From eq.\ (\ref{Hp}) with the above values of couplings it is easy to
see that the vacancy lines consisting of more then four members have
negative energy which caused their polymerization at an early stage of
evolution.

With the advent of time these polymers start to branch and cross-link
producing den\-dri\-mer-like structures (fig.\ \ref{ fig1}b).

\begin{figure} 
\includegraphics[viewport = -160 100 230 425, scale = 0.45]{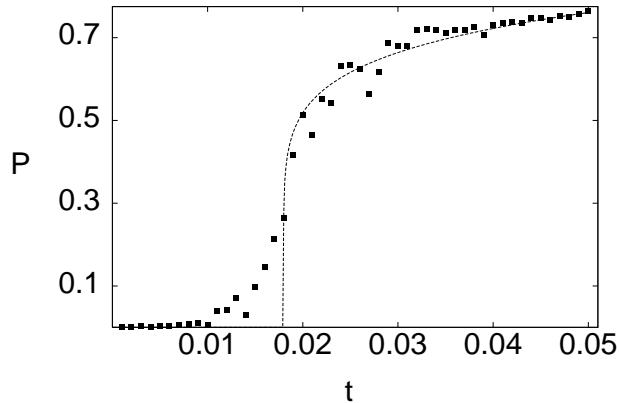}
\caption{Percolation order parameter P (black squares) as a function
of time elapsed since the initial random deposition; the dashed line is
a least square fit to the universal critical dependence 
\mbox{$\mbox{Const}(t-t_c)^\beta$} {\cite{stauffer}}.}
\label{fig2}
\end{figure}

Then, at some finite time $t_c$ the kinetic gelation of these polymers
takes place. In fig.\ \ref{fig2} the probability of a vacancy to belong
to the gel fraction is shown together with the best fit to the critical
curve of 2D percolation \cite{stauffer,gel1}. In simulations the gel
fraction was calculated in the standard way \cite{stauffer,gel1,gel2}
for two sizes of the system $L =$ 320 and 500 and then interpolated to
$N = \infty$.  As we see, the agreement of the MC data with the
theoretical values is not quite good but the problem of irreversible
gelation is known to be difficult for simulation \cite{gel1,gel2}, so
we invoke the notion of universality and the exactly known critical
exponent $\beta = 5/18$ \cite{stauffer,gel2} to retrieve the value of
$t_c\approx0.018$ from our data.

\begin{figure} 
\includegraphics[viewport = -80 0 230 180, scale = 1.1]{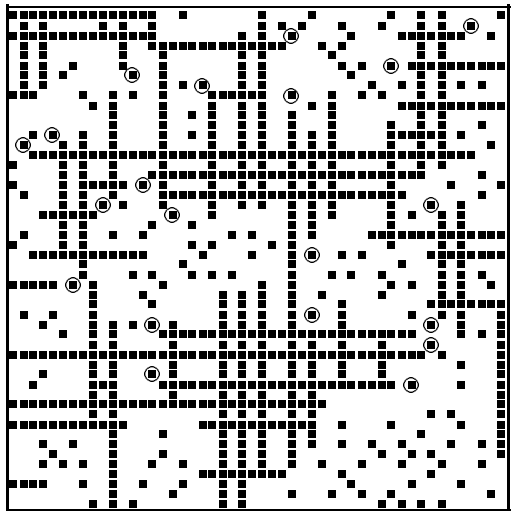} 
\caption{Typical local structure of the gel phase at $\beta = 6$ and
$t=0.05$; vacancies are shown by the black squares (only a $50\times50$
block of the $500\times500$ simulation box is shown). Enclosed in
circles are the ``active'' vacancies which at the next hop can become
attached to the nearby polymer chain.}
\label{fig3}
\end{figure}

In fig.\ \ref{fig3} is depicted a typical configuration of vacancies in
the gel phase. Noticeable is the cage structure characteristic of the
glass state \cite{angell}. Furthermore, the structure seen in this
figure is very close to a local energy minimum. A careful inspection
shows that any allowed hop of a vacancy belonging to the gel fraction
will lead to the energy growth while the movements of those that do not
belong to the gel leave the energy constant in the majority of cases.
Some movements of the free vacancies also augment the energy and only
the 20 ``active'' ones which we enclosed in the circles can diminish
the energy by attaching themselves to a nereby polymer chain. Thus, the
system is quite close to a minimum of the energy
landscape~\cite{angell,ritort} but is not at the bottom of it. On the
other hand, one can find in fig.\ \ref{fig3} $N_a\approx60$ ``active''
polymer ends which we define as those with the smallest detachment
energy $\Delta E_{\rm min} = -(W+V)$ [see eq.\ (\ref{Hp})]. Becase the
vacancy as a rule has 3 neighboring sites to hop, the number of
detached monomers can be estimated as
\[
N_d  = 3N_a\exp\left(-\frac{\Delta E_{\rm min}}{k_BT}\right)
= 3N_a\exp\left(\frac{W+V}{k_BT}\right).
\]
Substituting $(W+V)/k_BT=-2$ from eq.\ (\ref{ 2}) we get 
\[
N_d\approx 24.
\]
Thus, the number of active monomers approximately corresponds to its
equilibrium value which means that the system is close to a minimum of
the free energy $F$. This explains its relative stability because
the thermodynamic driving force proportional to $\delta F$ is close to
zero.

\begin{figure} 
\includegraphics[viewport = -120 130 230 500, scale = 0.45]{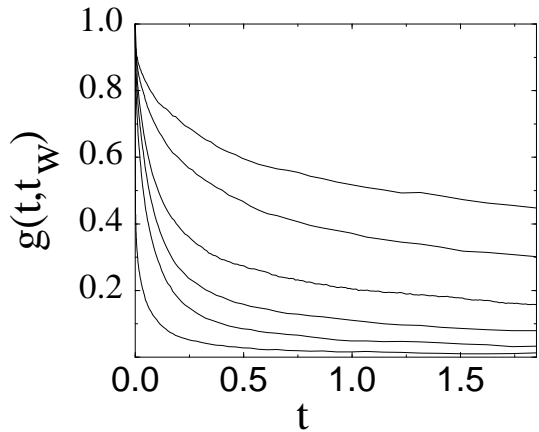}
\caption{Autocorrelation function eq.\ (\ref{g}) shown (from bottom up) 
for $t_{\mbox{w}} =$ 0, 0.01, 0.05, 0.1, 0.25, and 0.5.}
\label{ fig4}
\end{figure}

The local autocorrelation function 
\begin{equation}
\label{g}
g(t,t_{\mbox{w}}) = 
\frac{1}{Nc(1-c)}\sum_{\bf i}[\left<{n}_{\bf i}(t+t_{\mbox{w}})
{n}_{\bf i}(t_{\mbox{w}})\right>-c^2]
\end{equation}
is plotted in fig.\ \ref{ fig4} for several values of the aging time
$t_{\mbox{w}}$.  Because of much larger time intervals, the simulations
were performed on systems of size $L=50$ with averaging over 100
realizations.  At equilibrium $g(t,t_{\mbox{w}})$ should be
$t_{\mbox{w}}$-independent while in our case relaxation considerably
slows down with age.  Noticeable also is the separation of time scales
characteristic of glassy state: quick relaxation at early times and
much slower one at later stages of evolution (the so-called $\beta$-
and $\alpha$-processes
\cite{angell}).

Further aging leads to the compactification of the gel fraction which
at its late stages leads to a transient ordered
state~\cite{khachaturyan} (fig.\ \ref{ fig1}c). Finally, the system
arrives at its equilibrium phase separated state which can be
interpreted as the glass crystallization (fig.\ \ref{ fig1}d).

Kinetic glassy phenomena are known to be extremely sensitive to such
parameters as temperature and the cooling rate. For example, the
so-called escape time $\tau_{out}$ \cite{angell} corresponding roughly
to half the time needed for the glass to crystallize (in our case for
the atoms and vacancies to separate) was in our simulations with the
instant quench approximately two orders of magnitude greater than
$t_c$. Under a step annealing, on the other hand, we were able to
achieve $\tau_{out} = O(t_c)$.

Thus, we have shown that the model with the hamiltonian (\ref{Hp}) exhibits
a glass transition of the type of the kinetic gelation of lattice
polymers~\cite{stauffer,gel1,gel2}. The transition is of the second
order, so we may hope that the universality hypothesis apply. This
means that in the vicinity of the critical point $t_c$ the behaviour is
insensitive to the microscopics of the system and should be the same as
in the off-lattice case. Furthermore, if the universality class of the
glass transition is unique, than the universal behaviour should be the
same in all systems mentioned at the beginning of this paper.  This
seems to be too daring a conclusion because physically polymers and, {\em e.
g.},  metallic glasses look extremely disparate. However, off-lattice
numerical simulations of refs. \cite{nature2,donati} showed that indeed
in the vicinity of the glass transition both systems look very much
alike.  Moreover, even at the microscopic level the real space
Adam-Gibbs~\cite{gibbs} (see also ref.~\cite{nature3} and references
therein) cooperative structures in an off-lattice Lennard-Jones liquid
look similar to the polymers of fig.\ \ref{ fig1}b. This is yet another
argument in favor of the hypothesis that the proposed model may really
describe a universality class of the glass transition in 2D liquids.
\acknowledgements
We are grateful to O. Bengone for his help in optimization of the MC
code.  Financial support of CNRS through collaborative grant is
acknowledged. One of us (V. T.) expresses his gratitude to University
Louis Pasteur de Strasbourg and IPCMS for their hospitality. IPCMS is
UMR ULP-CNRS 7504.

\end{document}